\documentclass[journal,12pt,onecolumn,draftclsnofoot]{IEEEtran}
\usepackage{cite}
\usepackage{xspace}
\usepackage{caption}
\usepackage{subcaption}
\usepackage{xcolor}
\usepackage[cmex10]{amsmath}
\usepackage{mathtools,amssymb,lipsum}
\usepackage{lettrine}
\usepackage{cite}
\usepackage{bm}

\usepackage{algorithm,algorithmic}
\makeatletter
\newcommand{\multilines}[1]{%
	\begin{tabularx}{\dimexpr\linewidth-\ALG@thistlm}[t]{@{}X@{}}
		#1
	\end{tabularx}
}
\makeatother
\usepackage{array}
\usepackage{url}  
\usepackage{amsthm}
\usepackage{multirow}
\usepackage{color}
\usepackage{endnotes}
\usepackage{framed}
\usepackage{cool} 
\usepackage{enumerate} 
\usepackage{url}
\usepackage[subnum]{cases}
\usepackage[nocomma]{optidef}
\usepackage{etoolbox}

\usepackage{graphicx}
\DeclareGraphicsExtensions{.pdf,.eps}

\makeatletter
\newcommand{\multiline}[1]{%
  \begin{tabularx}{\dimexpr\linewidth-\ALG@thistlm}[t]{@{}X@{}}
    #1
  \end{tabularx}
}

\makeatletter
\patchcmd{\@maketitle}
{\addvspace{0.5\baselineskip}\egroup}
{\addvspace{-2\baselineskip}\egroup}
{}
{}
\makeatother

\begin{document}

\title{Latency-sensitive Service Delivery with UAV-Assisted 5G Networks}


\author{Shashi~Raj~Pandey\thanks{Shashi Raj Pandey, K. Kim, M. Alsenwi, Y. K. Tun, and C.S Hong are with the Department of Computer Science and Engineering, Kyung Hee University, Yongin 17104, South Korea (email: shashiraj@khu.ac.kr, glideslope@khu.ac.kr, malsenwi@khu.ac.kr, ykyawtun7@khu.ac.kr, cshong@khu.ac.kr).},~\IEEEmembership{Student Member,~IEEE,}~Kitae~Kim,~Madyan~Alsenwi,\\~Yan~Kyaw~Tun,~Zhu~Han\thanks{Z. Han is with the Electrical and Computer Engineering Department, University of Houston, Houston, TX 77004, USA, and also with the Department of Computer Science and Engineering, Kyung Hee University, Yongin 17104, South Korea (email: zhan2@uh.edu).},~\IEEEmembership{Fellow,~IEEE,} ~Choong~Seon~Hong,~\IEEEmembership{Senior Member,~IEEE} }


\maketitle
\begin{abstract}
In this letter, a novel framework to deliver critical spread out URLLC services deploying unmanned aerial vehicles (UAVs) in an out-of-coverage area is developed. To this end, the resource optimization problem, i.e., resource blocks (RBs) and power allocation, and optimal UAV deployment strategy are studied for UAV-assisted 5G networks  to jointly maximize the average sum-rate and minimize the transmit power of UAV while satisfying the URLLC requirements. To cope with the sporadic URLLC traffic problem, an efficient online URLLC traffic prediction model based on Gaussian Process Regression (GPR) is proposed which derives optimal URLLC scheduling and transmit power strategy. The formulated problem is revealed as a mixed-integer nonlinear programming (MINLP), which is solved following the introduced successive minimization algorithm. Finally, simulation results are provided to show our proposed solution approach's efficiency.

\end{abstract}


\IEEEpeerreviewmaketitle
\section{Introduction}

Recently, unmanned aerial vehicles (UAVs)-assisted cellular networks are deployed as a promising alternative to handle out-of-coverage issues. These UAVs work as a flying base station (BS) that offers computing and communication facilities for Internet of Things (IoT) devices and applications, e.g., disaster and rescue, autonomous control, military operations, and smart farming \cite{mozaffari2019tutorial, tun2020energy}. However, UAVs are energy-constrained, which needs to delicately allocate its available resources while concurrently serving as a local multi-access edge computing (MEC) infrastructure. Besides, to fully reap the benefits of UAVs, it is essential to have its efficient integration with the 5G New Radio (NR) standards to deliver latency-sensitive data packets.    

5G NR standardization has been a significant paradigm shift to realize full-fledged communication networks for the next-generation applications. The Third Generation Partnership Project (3GPP) Release 15 5G-NR \cite{3gpp} supports connectivity for massive device densities, high data rate, and ultra-reliable low-latency communication (URLLC) services. The key feature of URLLC supports a high level of reliability and low latency services, i.e., the latency of less than 1\textit{ms} while guaranteeing packet error rates (PER) in the \textit{order-of-five} (10$^{-5}$) \cite{3gpp}. This mandates immediate transmission of critical URLLC packets over the short transmission time interval (sTTI) \cite{3gpp} to meet the stringent latency-reliability requirements. Hence, recent studies focus on resource block\footnote{Resource block (RB) is the smallest unit of bandwidth resources defined in Long Term Evolution (LTE) \cite{3gpp}.} (RB) allocation problem to deliver URLLC services \cite{anand2020joint,bennis2018ultrareliable, alsenwi2019embb, huang2020deep, kasgari2019model, alsenwi2021intelligent}. However, the solution to overcome the challenges due to the dynamic nature of URLLC traffic is non-trivial. 
Inline with \cite{alsenwi2019embb}, several approaches have been recently proposed to optimize resource allocation \cite{anand2020joint, huang2020deep}, considering a general arrival process to capture dynamic URLLC traffic and perform RB allocation. Authors in \cite{kasgari2019model} proposed a model-free approach to guarantee end-to-end latency and end-to-end reliability imposing latency constraints in the optimization problem. However, the authors in  \cite{kasgari2019model, alsenwi2019embb, anand2020joint, huang2020deep} consider a typical network infrastructure with a fixed BS, having no energy restrictions, and the mobile users deployed randomly under its coverage area. Furthermore, the challenges of integrating 5G features on energy-constrained UAV systems, particularly serving out-of-coverage users, are still overlooked in recent studies \cite{mozaffari2019tutorial, khawaja2019survey, tun2020energy}. 

In this work, we first leverage the benefits of UAVs to ensure latency-sensitive data transmission and offer 5G services in an out-of-coverage area with unmanned aerial systems. In particular, we adopt a Gaussian Process Regression (GPR) \cite{williams1998prediction} approach to capture the network dynamics and predict the URLLC traffic online for executing an efficient resource allocation (i.e., RBs and power) and optimal deployment strategy of the UAV. GPR, a flexible and robust active learning approach that shows merit in tackling parametric models' issues \cite{williams1998prediction} and best-capture uncertainty in time-varying processes, allows UAVs to predict latency-sensitive data packets before serving the remote IoT devices. Hence, leveraging the methodological advantages of a GPR-based approach, which offers better functional performance and analytical properties, we can predict the dynamic URLLC traffic in an online fashion. Moreover, unlike existing prediction algorithms such as Long Term Short Term Mermory (LSTM), a GPR mechanism works well in small datasets and efficiently handles random components while making prediction.

In summary, the main contribution of this paper is a novel framework to deliver critical URLLC services in an out-of-coverage area deploying UAVs. In doing so, we develop a practical integration of 5G features with UAV networks and leverage GPR to appropriately characterize and predict the dynamic URLLC traffic online. We later fuse this prediction information to optimize the radio resources, i.e., RBs and transmit power, and deployment strategy of the UAV. Then, we formulate a joint optimization problem for maximizing the average sum-rate while minimizing the transmit power of UAV with the constraints to satisfy stringent URLLC requirements. The formulated problem is revealed as an MINLP, which is NP-hard due to the binary constraint. Hence, we relax the binary constraints and decompose the proposed problem into three sub-problems, which are solved using a low-complexity near-optimal successive minimization algorithm. Simulation results show that the proposed approach achieves a performance gain of up to 24.2\% as compared with the baseline for satisfying the reliability constraints.

To our best knowledge, this is the first work that adopts GPR for performing dynamic URLLC traffic prediction and resource optimization to guarantee maximum average sum-rate and minimum transmit power, jointly, in a UAV-assisted 5G network.
\begin{figure}[t!]
	\centering
	\captionsetup{justification = centering}
	\includegraphics[width=4.5in]{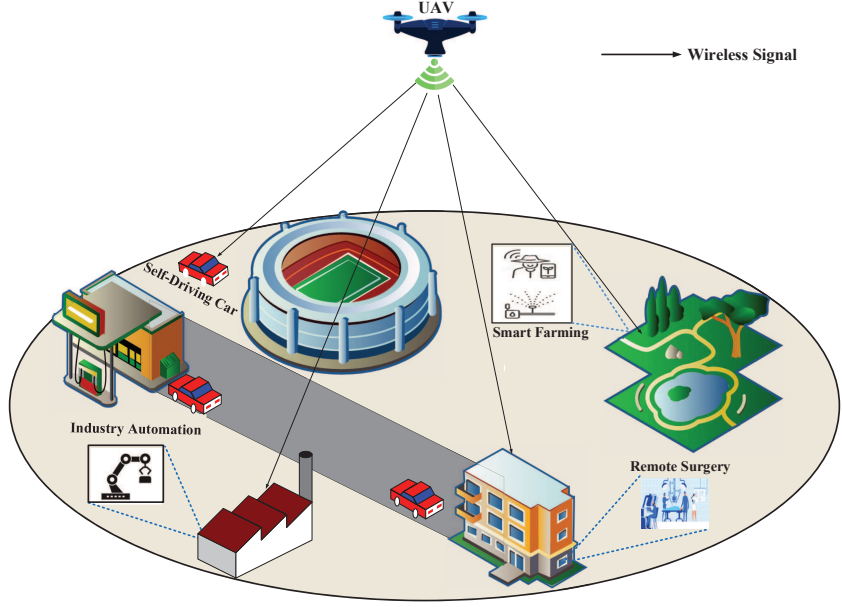}
	\caption{Illustration of our system model.}
	\label{fig:sys_model}
\end{figure}

\section{System Model and Problem Formulation}
\subsection{System Model}
In this work, as depicted in Fig.~1, we consider a wireless network system where a single UAV is deployed in an out-of-coverage area to provide wireless communication services to a set of URLLC users\footnote{We will use the term ``users" to denote URLLC users henceforth.} $\mathcal{U}$ of $|\mathcal{U}|=U$ at time slot in the set $\mathcal{T}$ of $|\mathcal{T}|=T$. We fixed the location of UAV at an altitude $H$, and the horizontal coordinates are initialized at $(x,y)$; thus, the position of UAV is $\boldsymbol{c}=[x, y, H]$. Similarly, the location of URLLC user is $\boldsymbol{o}_u=[x_u, y_u], \forall u \in \mathcal{U}$. The system available total bandwidth is divided into a set of RBs $\mathcal{B}$ of $|\mathcal{B}|=B$. We considered line-of-sight (LOS) link is available, and the orthogonal frequency division multiple access (OFDMA) scheme is adopted to share radio resources amongst the URLLC users. Let $a_{u}^{b} (t) \in \{0, 1\}$ be the RB assignment variable at time slot $t$ defined as
\begin{equation}
a_{u}^{b} (t) =
\begin{cases}
1, \; \; \text{ if user $u$ is assigned to RB $b$ at time slot $t$},\\
0, \; \; \text{otherwise}.
\end{cases}
\end{equation}
Next, considering a general channel fading model \cite{zhan2017energy}, we can define the channel coefficient between the user $u$ assigned to RB $b$ and UAV at time slot $t$ as $h_u^b(t)= \sqrt{\gamma_u^b (t)}\rho_u^b(t)$, where $\rho_u^b(t)$ is a small-scale fading coefficient and $\gamma_u^b (t)$ is the attenuation factor depending upon the distance between the user and UAV. Formally, we define $\gamma_u^b (t)$ as
\begin{align}
\gamma_u^b (t) &= \gamma_0 d_u^{-\theta},   \forall u\in \mathcal{U}, \forall b \in \mathcal{B}, \forall t \in \mathcal{T}, \label{SINR}
\end{align} 
where, respectively, $d_u$ is the distance between the UAV and user $u$ defined as $d_u = \sqrt { H^2 + || \boldsymbol{c} - \boldsymbol{o}_u||^2 }$, $\theta$ is the path loss exponent, and $\gamma_0$ is the channel power gain at the reference distance $d_0 =1$ m. Then, considering the LoS path, the small-scale fading can be modeled by the Rician fading with $\mathbb{E}|\rho_u^b(t)|^2 =1$ as
\begin{align} \label{SINR}
\rho_u^b(t) &=  \sqrt{\frac{\Tilde{K}_u^b(t)}{\Tilde{K}_u^b(t)+1}}\rho + \sqrt{\frac{1}{\Tilde{K}_u^b(t)+1}}\Tilde{\rho},
\end{align} 
where $\rho$ and $\Tilde{\rho}$, respectively, denote the deterministic LoS channel component with $|\rho| =1$ and the random scattered component which is a symmetric complex Gaussian
(CSCG) random variable having zero mean and a unit variance, and $\Tilde{K}_u^b(t)$ is the Rician factor of the channel between the user $u$ assigned to RB $b$ and UAV at time slot $t$. Therefore, the achievable rate expression over the RB $b$ for user $u \in \mathcal{U}$ at time slot $t$ of a block fading channel in a finite-blocklength regime \cite{polyanskiy2010channel, she2021tutorial}, is defined as 
\begin{align} \label{SINR}
r_u^b (t) &=  a_{u}^{b} (t)\Bigg[\omega^b    \log_2\left(1 + \dfrac{p_u^b (t)|h_u^b (t)|^2}{n_0 \omega^b}\right) - \sqrt\frac{V_u^b(t)}{n_u^b(t)}Q^{-1}(\Theta)\Bigg], \forall u\in \mathcal{U}, \forall b \in \mathcal{B}, \forall t \in \mathcal{T},
\end{align} 
where $\omega^b$ is the bandwidth of each RB, $n_0$ is the additive white Gaussian noise power density, $p_u^b (t)$ is the transmit power of the UAV over RB $b$ at time slot $t$, respectively; $V_u^b(t)$ is the channel dispersion\footnote{It captures the stochastic variability of the channel of user $u$ at time slot $t$.} given by 	$V_u^b(t)=1-\frac{1}{\left(1+\frac{p_u^b(t)|h_u^b(t)|^2}{n_0\omega^b}\right)}$, $n_u^b(t)$ is blocklength, and $Q^{-1}(\Theta)$ is the Q-function with error probability $\Theta >0$, respectively.

Let $L_u(t)$ denotes the random URLLC traffic arrivals at time slot $t$ of user $u$. Therefore, the system reliability constraint can be formally defined as 
\begin{equation}
    \mathsf{Pr} \left[ \sum_{b=1}^{B}r_u^b(t) \leq \beta L_u(t)    \right]  \leq \epsilon , \forall u \in \mathcal{U}, \forall t \in \mathcal{T},
\end{equation}
where $\beta$ denotes the URLLC packet size, and $\epsilon$ is a small outage threshold value. Then, by using the Markov's Inequality \cite{alsenwi2019embb}, we can rewrite the reliability constraint in (4) as a linear constraint as follows:  
	\begin{equation}
	\mathsf{Pr} \left[ \sum_{b=1}^{B}  r_u^b (t)\leq \beta L_u(t)    \right]\leq \frac{\beta \mathbb{E}[L_u(t)]}{\sum_{b=1}^{B} r_u^b (t)}, \forall u \in 
	\mathcal{U}, \forall t \in \mathcal{T}.
	\label{eq:markov}
	\end{equation}

\subsection{Problem Formulation}
    In order to satisfy the stringent latency requirements of URLLC traffic \eqref{eq:markov}, we need to consider the limited resource capacity, i.e., transmit power and RBs, at the UAV and its optimal deployment strategy. Therefore, we formulate our optimization problem to jointly maximize the average sum-rate and minimize the transmit power of the UAV, while ensuring the URLLC constraints, as follows:
   	\begin{maxi!}[2]                
		{\boldsymbol{a, p, c}}                               
		{ \frac{1}{T}\sum_{t=1}^{T}\bigg( \sum_{u=1}^{U} \sum_{b=1}^{B} r_u^b (t) - \zeta\sum_{u=1}^{U} \sum_{b=1}^{B} p_u^b (t) \bigg)}{\label{opt:P1}}{\textbf{P:}}   
		\addConstraint{ \sum_{b=1}^{B}r_u^b (t)\geq \frac{\beta\mathbb{E}[L_u(t)]}{\epsilon},}{\; \; \forall u \in \mathcal{U}, \forall t \in \mathcal{T}, \label{cons1:reliablity}}
		\addConstraint{\sum_{u=1}^{U} a_{u}^{b} (t) \leq 1,}{\; \; \forall b \in \mathcal{B}, \forall t \in \mathcal{T}, \label{cons2:association}}
		\addConstraint{a_{u}^{b} (t) \in \{0,1\},}{\; \; \forall u \in \mathcal{U}, b \in \mathcal{B}, \forall t \in \mathcal{T}, \label{con3:association_variable}}
		\addConstraint{\sum_{u=1}^{U} \sum_{b=1}^{B} a_{u}^{b} (t) p_{u}^{b} (t)\leq P^{\mathsf{max}}, \forall t \in \mathcal{T},}{ \label{con4:powerbudget}}
		\addConstraint{0 \leq  p_{u}^{b} (t) \leq  P^{\mathsf{max}},}{\forall u \in \mathcal{U}, \forall b \in \mathcal{B}, \forall t \in \mathcal{T}, \label{con5:powerbudget_variable}}	
	\end{maxi!}
where $\boldsymbol{a}$ and $\boldsymbol{p}$ are, respectively, the matrix of resource allocation and transmit power, $\boldsymbol{c}$ is the location of the UAV, and $\zeta > 0$ is a scaling constant. Here, $\boldsymbol{a}$ characterizes the mapping between the RBs in $\mathcal{B}$ and the number of users $\mathcal{U}$. \eqref{cons1:reliablity} is the URLLC reliability constraint, and constraints \eqref{cons2:association} and \eqref{con3:association_variable} ensure one RB is assigned to only one user at most. Constraints \eqref{con4:powerbudget} and \eqref{con5:powerbudget_variable} define the total transmit power of UAV over all RBs is bounded by the system power budget $P^{\mathsf{max}}$. 

\section{Proposed Solution Approach}

The formulated problem in \eqref{opt:P1} is an MINLP, which may require exponential-complexity to solve. To solve \eqref{opt:P1} efficiently and sub-optimally, we decompose it into three sub-problems: (i) \textit{RB allocation}, (ii) \textit{transmit power allocation}, and (iii) \textit{UAV location optimization.}

\subsection{RB Allocation Problem for a Given Power Allocation and UAV Location}

For a given power allocation and UAV location, we can relax the binary constraint \eqref{con3:association_variable} and recast the integer programming problem \eqref{opt:P1} as a RB allocation problem. Then, the fractional
solution is rounded to get a solution to the original integer
problem following the threshold rounding technique described in \cite{alsenwi2021intelligent}. Hence, we pose \eqref{opt:P1} as 
   	\begin{maxi!}[2]                
		{\hat{\boldsymbol{a}}}                               
		{\frac{1}{T}\sum_{t=1}^{T}\sum_{u=1}^{U} \sum_{b=1}^{B} r_u^b (t) }{\label{opt:P2}}{\textbf{P1: }}   
		\addConstraint{\eqref{cons1:reliablity}, \eqref{cons2:association}, \eqref{con4:powerbudget},}
		\addConstraint{\hat{a}_{u}^{b} (t) \in [0,1],}{\; \; \forall u \in \mathcal{U}, b \in \mathcal{B}, \forall t \in \mathcal{T}. \label{con3_P2:association_variable}}
	\end{maxi!}	
The above problem is a maximization problem with concave objective and linear constraints; hence, a convex optimization problem which can be solved efficiently using ECOS solver in the CVXPY toolkit.

\subsection{Transmit Power Allocation Problem for a Given RB Allocation and UAV Location}

For a given RB allocation and UAV location, we can recast the integer programming problem \eqref{opt:P1} as a transmit power allocation problem as  
   	\begin{maxi!}[2]                
		{\boldsymbol{p}}                               
		{\frac{1}{T}\sum_{t=1}^{T} \bigg(\sum_{u=1}^{U} \sum_{b=1}^{B} r_u^b (t) - \zeta\sum_{u=1}^{U} \sum_{b=1}^{B} p_u^b (t) \bigg)}{\label{opt:P3}}{\textbf{P2: }}   
		\addConstraint{\eqref{cons1:reliablity}, \eqref{con4:powerbudget}, \eqref{con5:powerbudget_variable}.}
	\end{maxi!}	
For any given RB allocation, the above problem is a
convex optimization problem which can be solved efficiently by the UAV.	

\subsection{UAV Location Optimization for a Given Power and RB Allocation}
For a given RB allocation from \textbf{P1} and power allocation from \textbf{P2}, the location optimization problem can be formulated as follows:
   	\begin{maxi!}[2]                
		{\boldsymbol{c}}                               
		{\frac{1}{T}\sum_{t=1}^{T} \bigg(\sum_{u=1}^{U} \sum_{b=1}^{B} r_u^b (t) - \zeta\sum_{u=1}^{U} \sum_{b=1}^{B} p_u^b (t) \bigg)}{\label{opt:P4}}{\textbf{P3: }}   
		\addConstraint{ \sum_{b=1}^{B}r_u^b (t)\geq \frac{\beta\mathbb{E}[L_u(t)]}{\epsilon},}{\; \; \forall u \in \mathcal{U}, \forall t \in \mathcal{T}, \label{cons1_P4:reliablity}}
	\end{maxi!}
The formulated problem is a convex optimization problem in $x$ and $y$ which can be shown following sequence of deduction from the proofs given in [\citenum{xu2020joint}, Appendix]. In particular, the Hessian of the inverse of the objective function \textbf{P3} is equivalently shown as a positive semi-definite, i.e., convex, about the UAV location. This means, with the direct consequence of simple composition rule that preserves convexity, the objective function is concave and positive. Thus, we have the maximization of a concave function as a convex. Therefore, we can solve \textbf{P3} using  ECOS  solver in the CVXPY toolkit.

However, to solve \textbf{P1}, \textbf{P2}, and \textbf{P3}, we first need to efficiently predict the expected random URLLC $L_u(t), \forall u$ traffic load at time $t$. A naive approach is to quantify $L_u(t), \forall u$ as a random variable with some known distribution \cite{alsenwi2019embb}; however, it may result poor performance in making online scheduling decision for URLLC traffic placements. Hence, we resort to a GPR approach, which is a flexible and robust mechanism to capture the network dynamics and provide online URLLC traffic prediction with minimal errors. 

Algorithm 1 summarizes the algorithm to solve \textbf{P} which must converge due to the fact that the overall problem is multi-convex; and hence, solving convex sub-problems in an iterative manner ensures convergence \cite{tun2020energy, xu2020joint, alsenwi2021intelligent}.
    	\begin{algorithm}[t!]
        	\caption{\strut Iterative solution approach for the relaxed problem}
        	\label{alg:profit}
        	\begin{algorithmic}[1]
        		\STATE{\textbf{Initialization:} Set $k=0$ and initial solutions $(\boldsymbol{a}^{(0)} (t), \boldsymbol{p}^{(0)} (t), \boldsymbol{c}^{(0)} (t))$;}
        		\STATE Obtain URLLC traffic prediction $L_u(t)$ from \eqref{channel_prediction};
        		\REPEAT
        		
        		\STATE{Compute $\hat{\boldsymbol{a}}^{(k+1)}(t)$ from (P1) at given $ \boldsymbol{p}^k(t)$, $\boldsymbol{c}^{(k)}(t)$};
        		\STATE{Compute $\boldsymbol{p}^{(k+1)}(t)$ from (P2) at given $ \hat{\boldsymbol{a}}^{(k+1)}(t)$, $\boldsymbol{c}^{(k)}(t)$};
        		\STATE{Compute $\boldsymbol{c}^{(k+1)}(t)$ from (P3) at given $ \hat{\boldsymbol{a}}^{(k+1)}(t)$, $ \boldsymbol{p}^{(k+1)}(t)$};
        		\STATE{$k = k + 1$}; 
        		\UNTIL{objective function converges.} 
        		\STATE{Recover a binary solution $\boldsymbol{a}^{(k+1)}(t)$ from $\hat{\boldsymbol{a}}^{(k+1)}(t)$ using the threshold rounding technique \cite{alsenwi2021intelligent}.}
        		\STATE{Then, set $\big(\boldsymbol{a}^{(k+1)}(t)$, $\boldsymbol{p}^{(k+1)}(t), \boldsymbol{c}^{(k+1)}(t) \big)$ as the desired solutions}.
        	\end{algorithmic}
        	\label{Algorithm}
 
        \end{algorithm}

\subsection{GPR-based URLLC Traffic Prediction}
Our aim is to perform an online prediction for the incoming URLLC traffic gain of the next time slot $\hat{L}_u(t+1)$ at each time slot $t$. To achieve that, we update the learning parameters over a moving window. Let $N$ be the window size, i.e., the window composed of the last $N$ time slots, in the set $\mathcal{N}$ of $|\mathcal{N}|=N$. The model parameters are trained on the data, i.e., URLLC traffic, inside the window. Then, trained parameters are used to predict the URLLC traffic load of the next time slot.   
In this view, for a finite data set $(t_n, L_u(t_n)), \;\forall n\in\mathcal{N}$, a general GPR-based prediction model \cite{williams1998prediction} can be modified as 
\begin{equation}
\hat{L}_u(t+1)=f(L_u(t))+\varepsilon, \; \forall u\in\mathcal{U},
\end{equation}
where $f(\cdot)$ is the regression function modeled as a Gaussian process with the mean function set to zero when there is no prior knowledge, and $\varepsilon$ is a Gaussian distribution random variable with $\sigma^2_{\varepsilon}$ variance and zero mean that represents the independent noise,
with the kernel function $g(\cdot)$ defined as
\begin{equation}
\begin{split}
g\big(L(t-m), L(t-n), \boldsymbol{\theta}\big)=\exp\bigg(\frac{-1}{\theta_1}\sin^2\Big(\frac{\pi}{\theta_2}\big(L(t-m)-L(t-n)\big)\Big)\bigg),
\end{split}
\end{equation}
where $m, n\in\{0, 1, 2, \dots, N\}$, and $\boldsymbol{\theta} =[\theta_1, \theta_2]$ defines a vector of the lengths and period hyper-parameters, respectively. Accordingly, the URLLC traffic load prediction at time slot $t+1$ is given as
\begin{equation}
\hat{L}_u(t+1)=g^{\dagger}(t)\boldsymbol{G}^{-1}[L_u(t-N), L_u(t-N+1), \dots, L_u(t)],
\label{channel_prediction}
\end{equation}
where $\boldsymbol{G}=[g(t-m, t-n)]$, and $g(t)=[g(t, t-n)], \; \forall m, n\in\mathcal{N}$.  Moreover, the variance (uncertainty) on the predicted value is given by
\begin{equation}
\textsf{Var}\big(\hat{L}_u(t+1)\big)=g(t, t)-g^{\dagger}(t)\boldsymbol{G}^{-1}g(t).
\label{variance}
\end{equation}

The traffic prediction is obtained from \eqref{channel_prediction} and exploring highly uncertain traffic provides more insight.
\begin{figure*}[t!]
	\centering
	\begin{subfigure}{0.3\linewidth}
		\includegraphics[width=\linewidth]{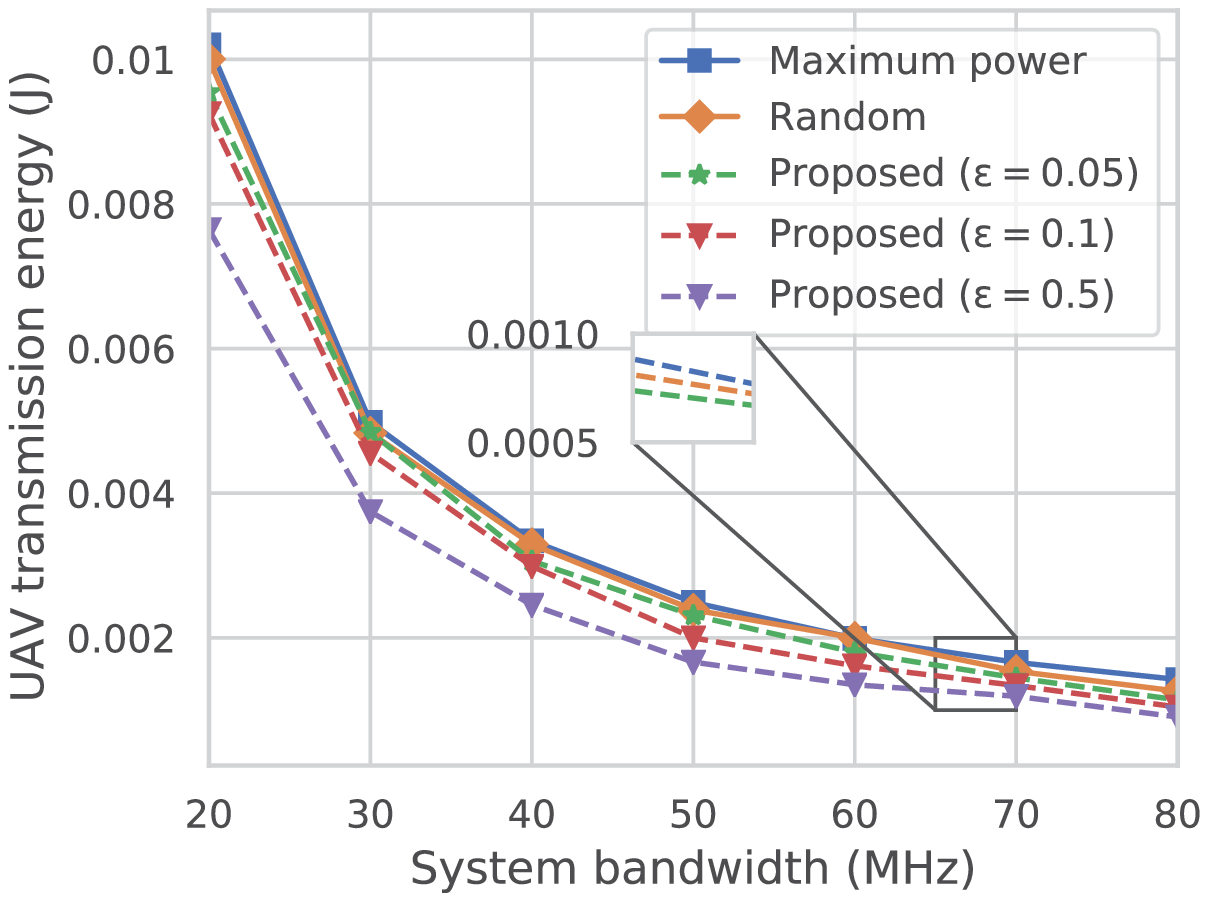}
		\caption{}
		\label{fig:a}
	\end{subfigure}
	\begin{subfigure}{0.3\linewidth}
		\includegraphics[width=\linewidth]{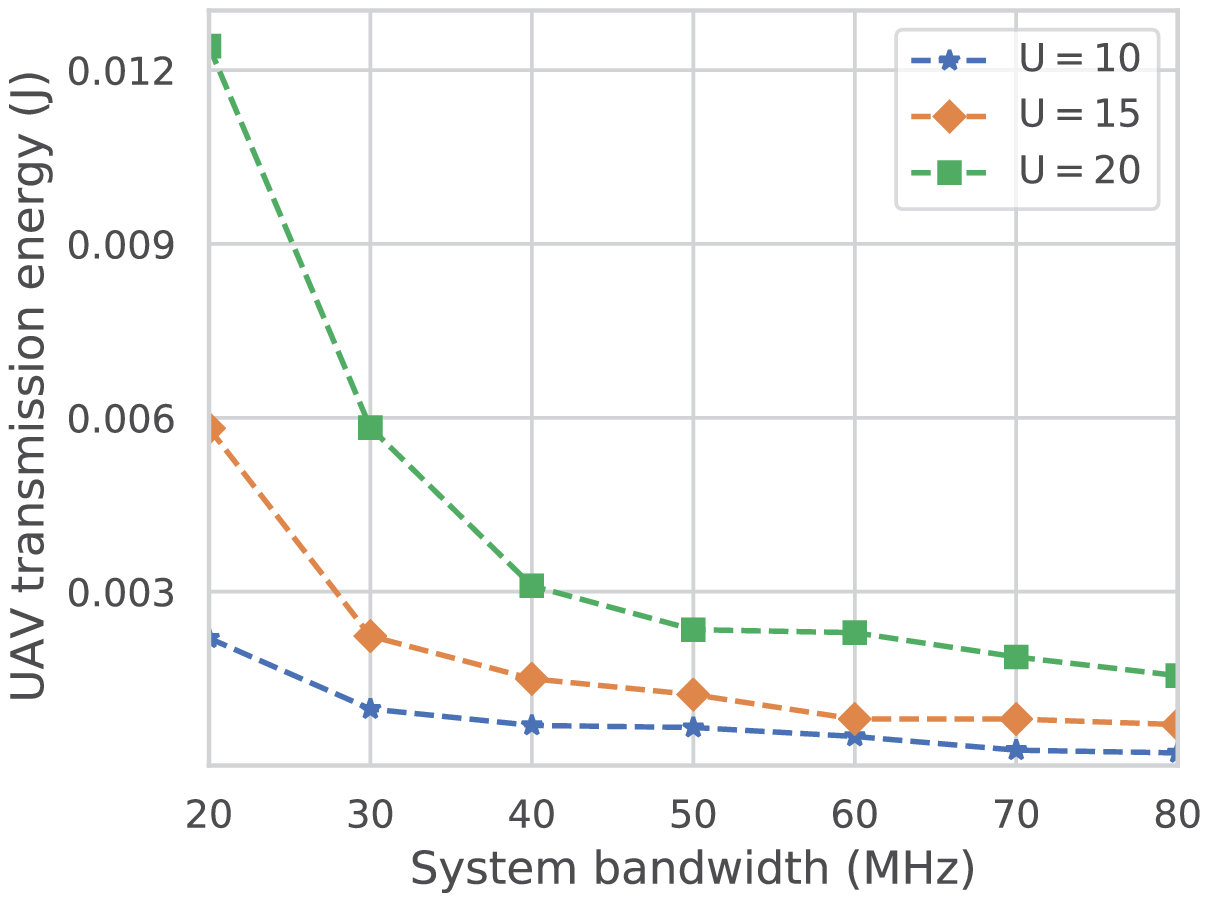}
		\caption{}
		\label{fig:b}
	\end{subfigure}
	\begin{subfigure}{0.3\linewidth}
		\includegraphics[width=\linewidth]{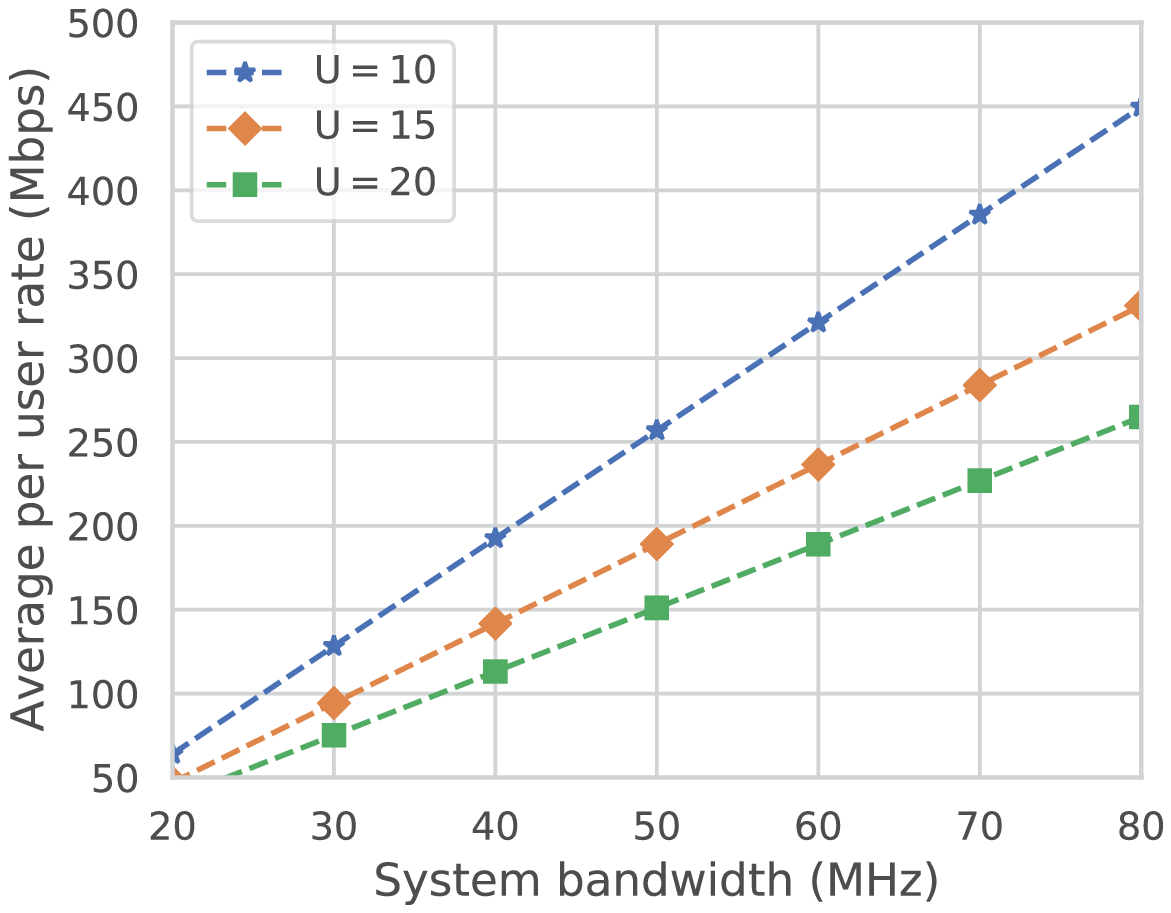}
		\caption{}
		\label{fig:rate}
	\end{subfigure}
	\caption{The tradeoff between UAV transmission energy and the system bandwidth varying (a) the outage threshold $\epsilon$, and (b) the network density $U$. Fig.~\eqref{fig:rate} shows the impact of the number of users on average per user data rate for different channel bandwidth.}
	\setlength{\belowcaptionskip}{-10pt}	
\end{figure*}

\section{Performance Evaluation}
In  our simulation, we consider the UAV is deployed at a fixed height in the range $H = [100, 150]$m with the coverage area of $(250\times250)$m$^2$. We set the number of users in the range $[5, 20]$, which are positioned randomly in the UAV's coverage area. We consider the bandwidth of each RB $\omega=180$ kHz. The total available transmit power is set as $P_{\max}=10$ Watts, the noise power density $n_0=-174$ dBm/Hz, and the channel gain at the reference distance is $\gamma_0 = -30$ dBm. We consider the URLLC packet size as $\beta=32$ bytes, and the window size $N=600$ time slots for traffic prediction. Due the absence of real URLLC traffic datasets, we adopt the real world stock market datasets\footnote{https://www.kaggle.com/szrlee/stock-time-series-20050101-to-20171231.} to replicate and characterize the URLLC traffic load dynamics; and hence, evaluating the performance of the proposed algorithms. Moreover, we show the performance evaluation of the proposed approach in terms of overall transmission energy that well-captures both the transmit power and average transmission rate.

Fig.~\eqref{fig:a} shows the impact of the outage threshold $\epsilon$ for different system bandwidth configuration on the overall UAV transmission energy with $U=20$. In this figure, decreasing $\epsilon$ leads to high URLLC reliability. We compare the performance of the proposed approach with two intuitive baselines \textbf{Maximum power}, which is the worst case scenario to satisfy reliability constraints, and \textbf{Random}, which considers random placement of the UAV. For a given system bandwidth, the UAV increases the transmit power, so higher transmission energy improves the average sum-rate required for obtaining higher reliability (i.e., smaller $\epsilon$); however, a performance gain of up to 24.2\% as compared with the Maximum Power and around 23\% with Random. On the other hand, we observe the UAV significantly lowers the transmit power when the available system bandwidth is high; thus, a low transmission energy. This is expected as a lower transmit power is sufficient enough to  maximize the average sum-rate for URLLC users that guarantees its reliability requirements. Thus, we observe the tradeoff between the overall transmission power and the average sum-rate, as defined in \eqref{opt:P1}. Moreover, the results are obtained after performing Monte-Carlo simulations to well-capture the variations in the UAV transmission energy when increasing the available system bandwidth. Such variations are particularly due to the random dynamics of wireless channel and uncertainty in the arrival of the URLLC traffic load.

Fig.~\eqref{fig:b} demonstrates the impact of network density, i.e., the number of URLLC users, on the UAV transmission energy for $\epsilon=0.1$. For a given available system bandwidth, it is shown that the UAV requires higher transmission energy to satisfy the stringent requirements of a large number of users. Moreover, with the increase in the available bandwidth, UAV can reduce its transmit energy consumption (low transmit power) without compromising the achievable average sum-rate. Similarly, Fig.~\eqref{fig:rate} evaluates the impact of the number of users on average per user data rate. In particular, we observe a sub-linear increase in the average per user rate with the availability of system bandwidth. Moreover, we also notice a negative impact of the network density on the per user rate, i.e., with the increasing number of users, the per user rate drops. This is intuitive as the radio resources will be shared amongst a larger number of users.

\begin{figure}[t!]
	\centering
	\includegraphics[width=0.4\linewidth]{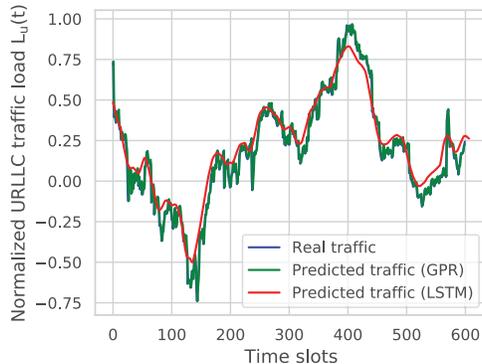}
	\caption{The performance of URLLC traffic load prediction model for N = 600.}
	\label{fig:gpr}
\end{figure}
Finally, Fig.~\ref{fig:gpr} captures the real trend of the normalized URLLC traffic load and the propagating uncertainty over time using a GPR prediction approach as compared to a two-layer LSTM network with ``\textit{rmsprop}" optimizer. In particular, it is observed that the prediction accuracy of GPR surpassed $99\%$ with a mean squared error (MSE) of 0.00288 when trained 600 data samples over the window of 600 time slots\footnote{In our formulation, we consider the worst-case scenario using chance constraint to ensure the reliability requirement of URLLC under such minimal errors.}, whereas an MSE of 0.015 with LSTM. Consequently, a better margin of performance gain is obtained while performing RB and transmit power optimization to satisfy the URLLC requirements.

\section{Conclusions}
In this letter, we have studied the problems of practical integration of 5G features with resource-constrained UAV to deliver URLLC services in an out-of-coverage area. In doing so, firstly, we have exploited a GPR approach to capture the real trend of URLLC traffic manipulating real world datasets. We have then formulated a joint optimization problem that incorporates optimal deployment strategy of UAV for maximizing the average sum-rate while minimizing its transmit power with the constraints to satisfy stringent URLLC requirements. We have revealed the formulated problem as an MINLP, challenging to solve directly using conventional optimization techniques. To tackle this issue, we have introduced the low-complexity near-optimal successive minimization algorithm. Finally, we have presented numerical results to validate the efficiency of our proposed solution approach where our approach outperforms the other baselines.

\ifCLASSOPTIONcaptionsoff
  \newpage
\fi

\vspace{-0.3cm}
\bibliographystyle{IEEEtran}
\bibliography{ref}
\end{document}